\documentclass[onecolumn,aps,pra,preprintnumbers,amsmath,amssymb,showkeys]{revtex4-2}

\usepackage{natbib}
\usepackage{graphicx}
\usepackage{bm}
\usepackage{color}

\usepackage{amsmath}
\usepackage{amsfonts}
\usepackage{amssymb}
\usepackage{makeidx}
\usepackage{graphicx}

\begin{document}
	
	\title{Decoherence dynamics   in a polaron system with collective dephasing}
	\author{Saima Bashir$^1$}
	
	\author{Muzaffar Qadir Lone$^2$ \footnote{corresponding author: lone.muzaffar@uok.edu.in}}
	\author{Prince A Ganaie$^3$}
	\affiliation{ $^1$ Department of Physics, National Institute of Technology, Srinagar-190006 India \\
		$^2$ Quantum Dynamics Lab, Department of Physics, University of Kashmir, Srinagar-190006 India}
	
	\begin{abstract}
		Within quantum information frameworks, managing decoherence stands as a pivotal task. The present work delves into decoherence dynamics of a { dressed qubit} , represented by a spinless fermion hopping between two lattice sites that are strongly coupled  to a collective bosonic bath. To simplify calculations under strong coupling, we adopt the Lang-Firsov transformation, effectively minimizing system-bath interactions. Within the polaron perspective using Ohmic bath spectral density with a Gaussian cutoff,  we identify a fundamental timescale $s$ (equivalently a length scale $l$), dictating coherence decay. Utilizing a quantum master equation in the energy eigen basis while maintaining fixed particle number, we demonstrate that coherence persists for small $s$ values but diminishes for larger ones. Additionally, we explore the utilization of $\pi$-pulses to manipulate decoherence within the system.
	\end{abstract}
	
	\maketitle
	\section{Introduction}
	
	Quantum mechanics has its foundations on the superposition principle that enables a particle to be simultaneously in many possible states\cite{1}. These superpositions lead to non-trivial correlations like entanglement that  are cornerstone to the quantum computation and quantum information processings\cite{2,3,4,5,6}. However, any real quantum system interacts with its environment (called as bath) \cite{7,8} and due to the extraordinary fragility of a quantum system, it often entangles itself quickly and strongly with a wide range of bath degrees of freedom. The resultant dissipative effects stemming from these  interactions cause the decoherence of quantum superpositions\cite{9,10,11}, and entanglement degradation resulting in the emergence of classical characteristics\cite{12}. Consequently, a significant obstacle to the development of quantum information processing systems, such as quantum computers, is decoherence\cite{13,14}. Thus, studying the dynamics of decoherence is crucial for harnessing the power of quantum mechanics for technological developments\cite{15,16,17}.
	
	In a given model, different time scales emerge due to system-bath couplings which govern the dynamics in a system under consideration. The interplay of these time scales give rise to Markovian and  non-Markovian dynamics\cite{18,19}. Typical time scales are bath correlation time, system relaxation time scale, etc. In this work, { we assume correlation time scale ($\tau_B$) of the bath to be much smaller than the relaxation time scale($\tau_R$) of the system ($\tau_R>>\tau_B$)}, a typical of Markovian approximation\cite{7}. Under this assumption,  the bath acts as the sink and information lost to the bath is not retrieved back during evolution\cite{19}. Furthermore, there has been intensive research on decoherence dynamics over last few decades \cite{20,21,22,23,24}and various definitions resulted to quantify decoherence in a given system\cite{25,26,27,28}. We simply consider the dynamics of off-diagonal elements of the density matrix in energy eigen basis of the underlying system to quantify decoherence. 
	
	In this paper, we consider a system described by a spinless fermion hopping between two lattice sites, and each site is assumed to have strong coupling with a collective bosonic bath\cite{29,30}. Such models can be realized via double well potential immersed in a Bose-Einstein condensate \cite{31,32,33,34,35,36}or by an impurity trapped into a given substrate\cite{37,38,39,40}. Next, we make perturbative calculations in the polaron frame (dressed basis), which is usually done via polaron transformation due to  Lang-Firsov\cite{41,42} to obtain an effective Hamiltonian where the  system-bath coupling gets  substantially reduced. { The Lang-Firsov transformation\cite{60,42}, also known as the polaron transformation, is a most effective mathematical technique used primarily in condensed matter physics to deal with strong electron-phonon interactions where the perturbative methods are no longer valid \cite{61}. This transformation utilizes the phonon basis residing at each lattice site to decouple the electron-phonon interaction in the strongly coupled electron-phonon system, for example, in Holestein Hubbard model. This transformation essentially captures the idea that the electron is surrounded by a phonon cloud, forming a polaron. This polaron acts like a composite particle, with greater effective mass and lower mobility compared to a free electron. The Lang-Firsov transformation allows us to model this behavior by modifying the hopping term, which is exponentially suppressed due to the electron-phonon coupling, enabling one to use perturbation theory as discussed below in section II.} Since, our main focus is on a detailed study of decoherence in this framework of  dressed (polaron) basis, we  use quantum master equation \cite{43,44,45} to examine the dynamic evolution of the density matrix elements in the single particle subspace of the two site system i.e $|T/S\rangle= \frac{1}{\sqrt{2}} [|10\rangle \pm |01\rangle]$, where $|0\rangle (|1\rangle)$ is an empty(filled) site.  Furthermore, we explore the decoherence protection schemes based on the $\pi$-pulses\cite{46,47,48,49,50,51}. We employ sequence of these pulses to control decoherence in the reduced system. 
	
	The outline of  rest of the paper follows as: In section II, we introduce our model and perform polaron transformation to obtain an effective Hamiltonian with reduced bandwidth. In section III, we employ quantum master equation under Markovian approximation to obtain the dynamics of the density operator of the system. We employ decoherence protection scheme in section IV. Finally, we conclude in section V.

	\section{Model Calculations}
	
	We consider a model in which a spinless fermion is hopping between two sites and  each site is strongly  coupled to a collective bosonic bath. Such models can be  realized in different ways. For example, we can realize this model via an impurity trapped in double well potential immersed in a Bose-Einstein condensate\cite{30} or by a double quantum dot etc\cite{38}. The total Hamiltonian of the system and bath can be written as
	\begin{eqnarray}
		\label{MM}
		H= H_S+ H_B + H_I,
	\end{eqnarray}
	where the system Hamiltonian is given by
	\begin{eqnarray}
		\label{M}
		H_S= \epsilon(a^{\dagger}_1 a_1 + a^{\dagger}_2 a_2) + J (a^{\dagger}_1a_2+ a^{\dagger}_2a_1).
	\end{eqnarray}
	Here, $\epsilon$ is the onsite energy, $J$ is the hopping strength. $a_{ p}, a_{ p}^{\dagger}$  $(p=1,2)$ are respectively  the annihilation and creation  operators for the fermions  at $p$th site and satisfy anti-commutation relation $\{a_{ p},a_j^{\dagger}\}=\delta_{{ p}j}$. The bath Hamiltonian $H_B= \sum_k \omega_k b^{\dagger}_k b_k$ with $\omega_k$ to be the energy of $k$th bath mode with  $b_k,b_k^{\dagger}$ its  annihilation and creation operators. Next we write interaction Hamiltonian as
	\begin{eqnarray}
		H_I= \sum_{{ p}=1}^2 \sum_k n_{ p} (g_{{ p}k} b_k + g^{\star}_{{ p}k} b^{\dagger}_k),
	\end{eqnarray}
	where $g_{pk}$ represents the  coupling strength at $p$th site and $n_p= a^{\dagger}_pa_p$. Since there exist strong strong coupling between the system and bath, therefore we transform to a polaron frame for the perturbative treatment of the problem.  We define an operator 
	$S=-\sum_{{ p},k}n_l\bigg(\frac{g_{{ p}k} }{\omega_k}b_k-\frac{g_{{ p}k}^*}{\omega_k }b_k^\dagger \bigg)$, so that in the  transformed frame, the total Hamiltonian can be written as (see appendix A for complete details);
	\begin{eqnarray}
		H' &=& e^S H e^{-S}= H_S^{\prime}+ H_{B}^{\prime}+ H_I^{\prime},
	\end{eqnarray}
	where 
	\begin{eqnarray}
		H_S^{\prime}= \tilde{J}[a_1^\dagger a_2+a_2^\dagger a_1]+ \epsilon(n_1+n_2)+V_{12}n_1n_2
	\end{eqnarray}
	represents the system Hamiltonian.
	The bath Hamiltonian is given by $H_B^{\prime}= \sum_k\omega_k b_k^\dagger b_k $, and the interaction Hamiltonian in the polaron frame is given by
	\begin{eqnarray}
		\label{inte}
		H_I^{\prime}=	\tilde{J}[\mathcal{B}a_1^\dagger a_2+\mathcal{B}^\dagger a_2^\dagger a_1].
	\end{eqnarray}
	Here, $	\tilde{J}=Je^{-\frac{1}{2}\sum_k |\alpha_k|^2}${   with $\alpha_k= \frac{g_{1k}-g_{2k}}{\omega_k}$ } is the effective hopping energy while
	$	\mathcal{B}=e^{\sum_{k}\alpha_k^*b_k^\dagger}e^{-\sum_{k}\alpha_k b_k}-1$ represents the transformed bath operators. 
	The collective bath mediates an interaction of strength  $ V_{12}=\sum_{k}\frac{g_{1k}^*g_{2k} + g_{1k}g_{2k}^*}{\omega_k}$ between the two sites. In a large of number of lattice sites e.g. a two dimensional lattice, this interaction is highly non-local. 
	
	Next, we evaluate the modified hopping rate $\tilde{J}$ using $g_{pk}= g_k e^{-i\vec{k}.\vec{r}}$ and the spectral density of the form $|g(\omega)|^2 =\alpha \omega e^{-\frac{\omega^2}{\Omega^2}}$, $\alpha$ is the intrinsic coupling constant\cite{52,53}. Let $\vec{l}$ be the distance of separation of the lattice sites, and using the linear dispersion $\omega=v k$, where $v$ is the typical speed of phonons, we write 
	\begin{eqnarray}
		\tilde{J}&=&J \exp\Bigg[-\frac{1}{2}\sum_k \frac{|g_{1k}-g_{2k}|^2}{\omega_k^2}\Bigg]=J\exp\Bigg[-\sum_k \frac{|g_k|^2 }{\omega_k^2}(1-\cos(\vec{k}.\vec{l})) \Bigg] \nonumber\\
		&=&{ J \exp\Bigg[-\frac{1}{(2\pi)^3}\int_{0}^{2\pi}d\phi\int_{0}^{\pi}\sin\theta d\theta\int_{0}^{\infty}k^2 dk \frac{|g(k)|^2}{\omega^2(k)}(1-\cos{(kl\cos\theta)}\cos\omega_{k}(t-\tau))\Bigg]}
		\nonumber \\ &=& J \exp\Bigg[- \frac{4\pi \alpha }{v^3} \int_0^{\infty} \!\!\!\!\!d\omega~ \omega e^{-\frac{\omega^2}{\Omega^2}} (1-\frac{\sin \omega s}{
			\omega s})\bigg] =J\exp\Bigg[- \frac{4\pi \alpha \Omega^2}{v^3} [\frac{1}{2}-\frac{F[\Omega s ]}{\Omega s}]\Bigg],
	\end{eqnarray}
	where in the second line, we have integrated over the solid angle  { $\theta$ and $\phi$ in the k-space} and $s= l/v$ is the intrinsic time scale in the system. {  $F[z] =\int_{0}^{\infty} dt e^{-t^2}{\sin zt}$ is the sine-transformed Dawson function. $\tilde{J}$ represents the reduced hoping energy, and is a smooth function  of the parameter $s$ and interaction coupling  $\alpha$, as shown in figures \ref{F}(a) and \ref{F}(b) . We see that that the effective coupling $\tilde{J}$ is substantially reduced. 
		
		\begin{figure}[t]
		\includegraphics[width=5.5cm,height=5cm]{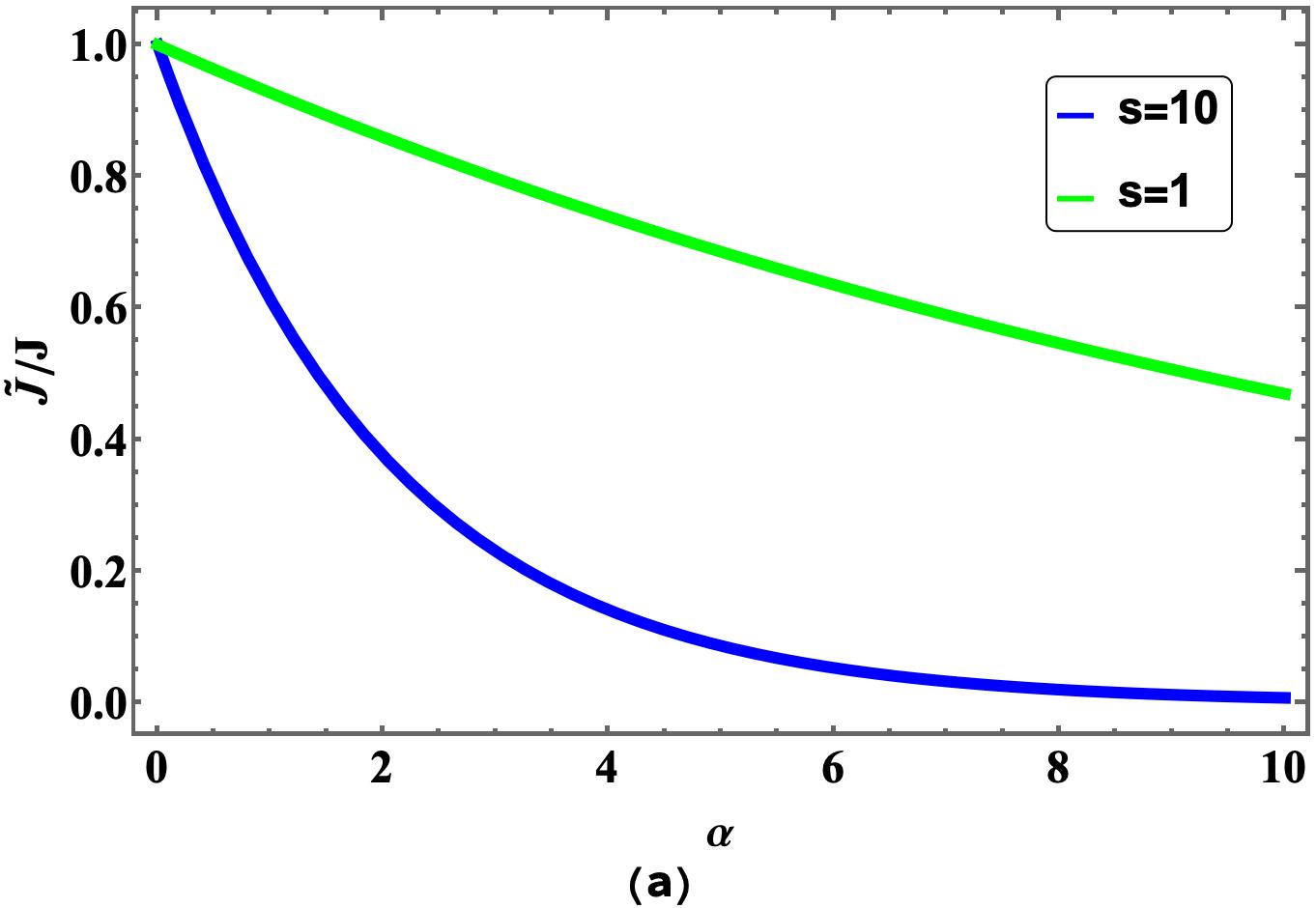} \hspace{0.31 cm}
		\includegraphics[width=5.5cm,height=5cm]{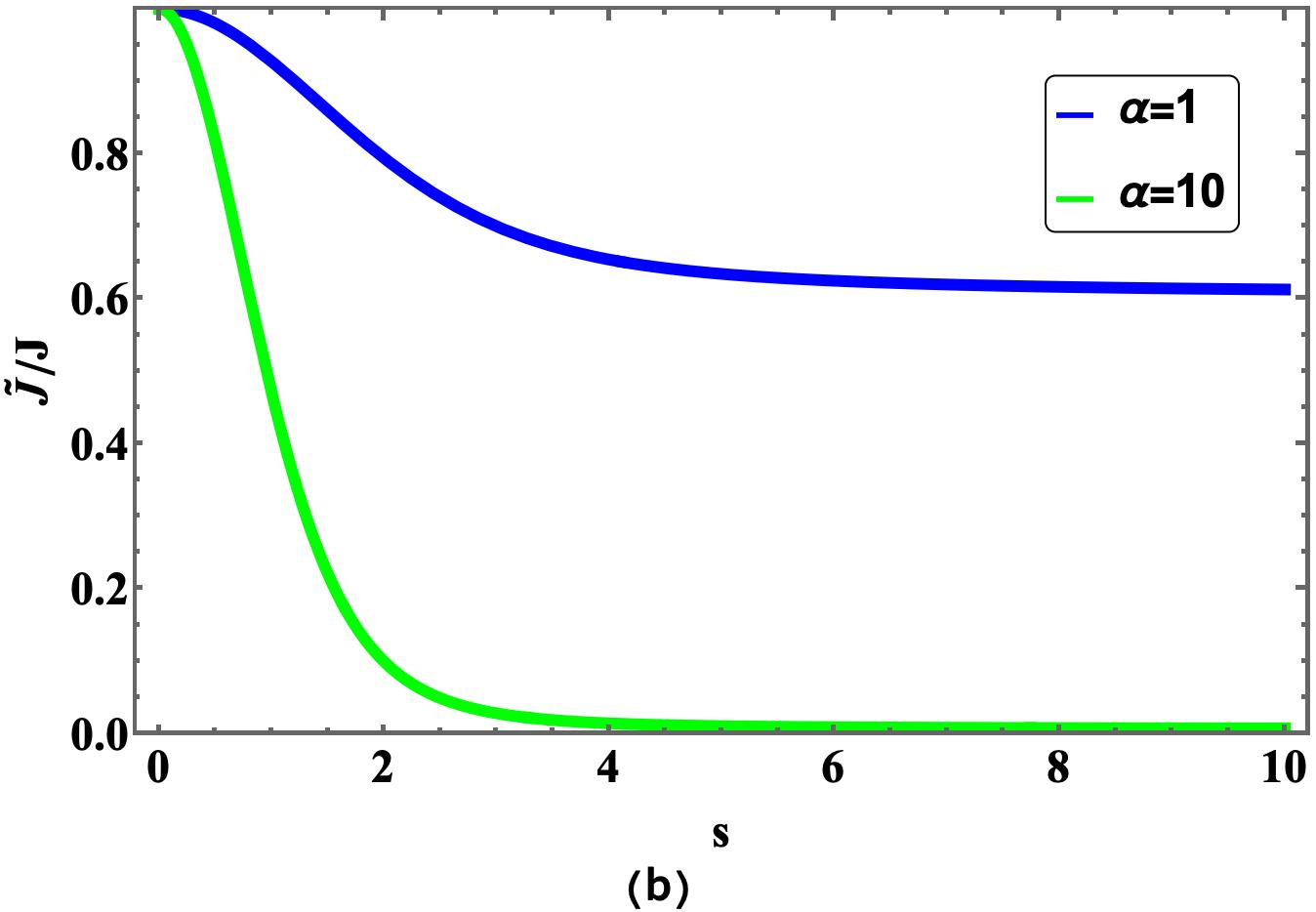} 
		\caption{ Here  we plot the $\frac{\tilde{J}}{J}$ with respect to s and bare coupling $\alpha$. (a) Represents the variation of $\frac{\tilde{J}}{J}$ with respect to bare coupling $\alpha$  for different values of s while (b) is the variation with respect to s for different  $\alpha$-values. In both cases, we see that the effective coupling decreases substantially in the polaron frame.}
			\label{F}
	\end{figure}
	Next, we look at the relevant parameters involved in the dynamics of the model considered in equation \ref{MM}.We have mainly  two time scales in our system, the adiabaticity parameter given by $\frac{J}{ E_B}$,where $E_B$ is  the energy scale set by the bath modes. Second, is the interaction scale set by $\frac{\alpha}{ E_B}$. If $\frac{J}{ E_B}<<1$, the dynamics is anti-adiabatic. Now in the polaron frame, the interaction energy scale changes to $\tilde{J}=Je^{-\frac{1}{2}\sum_k|\alpha_{k}|^2}$  as defined in equation \ref{inte} . Therefore, in the polaron frame the effective coupling strength is given by $\tilde{J}$ and  adiabaticity parameter changes to $\frac{\tilde{J}}{ E_B}$. The anti-adiabatic condition is reduced to $\frac{\tilde{J}}{ E_B}<<1$. While the Markovian approximation implies $\tau_R>>\tau_B$ that is $\frac{\tilde{J}}{\Delta E_B}<<1$. Thus anti-adiabatic condition corresponds to Markovian approximation.
	 Intuitively, this can be understood in the following way \cite{62,63,64}:
 \begin{figure}[t]
	\includegraphics[width=10cm,height=5cm]{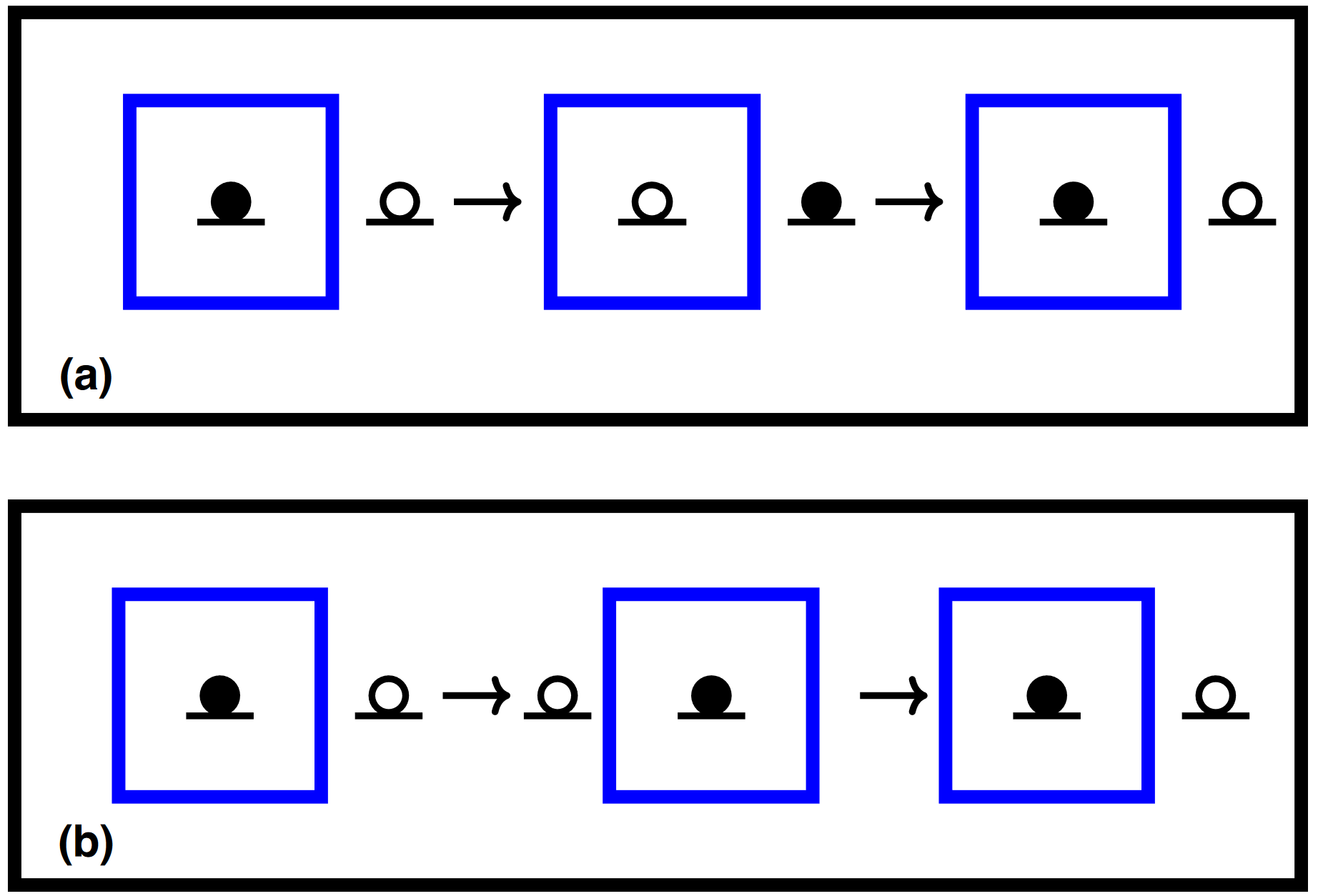}
	\caption{At second order of perturbation we have many processes, out which we have considered only these two, for the sake of completeness. The left is initial state, right the final while middle one represents the intermediate state. An empty circle has no particle while it is  present in the filled circle. Squares represent the lattice distortion. (a) represents the particle hoping without creating any additional lattice distortion while in (b) particle tunnels along with the phonon cloud.}
	\label{feyn}
\end{figure}
Since the model considered in this paper, is  a two site problem with one particle hopping between two sites while the phonons can be considered as some lattice distortion. In the transformed frame, we have a particle dressed with phonons-a polaron hopping with the effective rate $\frac{1}{\tilde{J}}$. Therefore, we can have processes where there is a full lattice distortion or relaxation that occur at the rate  $\frac{1}{\tilde{J}}$ and some other processes with negligible lattice distortion at the rate  $\frac{1}{J}$ as shown in the figure \ref{feyn}. These are two processes out of many at second order perturbation. The boxes represent the phonon modes (lattice distortion) with energy $\sim -\Delta E_B$ ($+\Delta E_B$)if particle is present (absent). Filled and empty circles correspond to a particle and no particle at a given site. Next, in the processes shown in the fig \ref{feyn}(a), the particle tunnels back to the original position through some intermediate states as allowed at second order of perturbation. In the intermediate state, the particle moves to second site leaving the first site with full lattice distortion and then finally going to original site without creating any new distortion. Thus during this process we conclude all these states have the same lattice distortion. This process occurs therefore at the rate $\tau_{Ra}^{-1}= \frac{J}{\Delta E_B}\times \tilde{J}= \frac{J^2 e^{-\sum_k|\alpha_k|^2}}{\Delta E_B}$. Now, in the part (b) of the fig \ref{feyn}, we can have a process where the particle hops as a polaron from one site to other and then back. This occurs at the rate $\tau_{Rb}^{-1}= \frac{\tilde{J}}{\Delta E_B}\times\frac{\tilde{J}}{\Delta E_B}=\frac{J^2e^{-2\sum_k|\alpha_k|^2}}{\Delta E^2_B} $. Thus the relative rate of these process are 
	$\frac{\tau_{Rb}^{-1}}{\tau_{Ra}^{-1}} \sim O(\frac{\tilde{J}}{\Delta E_B}) $.  Thus in the limit $\frac{\tilde{J}}{\Delta E_B}<<1$, there are several processes which are not resolved at this time scale by the master equation \ref{mast}.       
	Therefore in the Markovian approximation, system has no influence on the bath as  different processes do not contribute to the dynamics at this time scale.}
	
	\section{Decoherence Analysis}
	
	In this section, we use  master equation to analyze decoherence in the system. In the interaction picture, the master equation for the reduced density matrix of the system  is given by\cite{7,43,44,45}
	\begin{eqnarray}
		\label{mast}
		\frac{d \rho_s(t)}{dt} =-i {\rm Tr_B}[H'_I(t),\rho_s(0) \rho_B]-\int_0^t { d\tau}{\rm Tr_B}[	H_I^{\prime}(t),[	H_I^{\prime}(\tau),\rho_s(t)\rho_B]],
	\end{eqnarray}
	where $	H_I^{\prime}(t) = e^{i H_0 t} 	H_I^{\prime} e^{-i H_0 t}$ is the interaction Hamiltonian in the interaction picture with respect to $H_0 = 	H_S^{\prime} + 	H_B^{\prime}$. $\rho_B=\frac{e^{-\beta H_B^{\prime}}}{Z_b}$ is the bath density matrix, $Z_B$ is the partition function of the bath. Now, in order to simplify the above master equation for our problem, we first calculate the time dependence of interaction Hamiltonian $H_I^{\prime}(t)$ in interaction picture. Let $\{|E^f_{ p}\rangle \}$ define the energy eigen states of the system, $\{|\{m_k\}\rangle\}$ be the energy eigen states of the bath, then we write
	
	\begin{eqnarray}
	\label{LFH}
		H_I^{\prime}(t) &= &e^{i H_0 t} 	H_I^{\prime} e^{-i H_0 t}\nonumber \\
	& =&\sum_i  |E^f_{ p}\rangle \langle E^f_{ p}| \sum_{{ \{m_k\}}} |{ \{m_k\}}\rangle \langle { \{m_k\}}| e^{i({H}^{\prime}_S+{H}^{\prime}_B)t }{H}^{\prime}_I e^{{ -}i({H}^{\prime}_S+{H}^{\prime}_B)t } \sum_j  |E^f_j\rangle \langle E^f_j| \sum_{{ \{n_k\}}} |{ \{n_k\}}\rangle \langle { \{n_k\}}|  \nonumber \\
	&=& \sum_{i,j} \sum_{{ \{m_k\}},{ \{n_k\}}} e^{-i[(E^f_{ p}-E^f_j) + (\omega_{m_k}-\omega_{n_k})]t} |E^f_{ p} \rangle \langle E^f_j| |{ \{m_k\}}\rangle \langle { \{n_k\}}| \langle E^f_{ p}| \langle { \{m_k\}}|{ H}^{\prime}_I |E^f_j\rangle |{ \{n_k\}}\rangle.
\end{eqnarray}
We see that $\Delta E^f \equiv E^f_{ p}-E^f_j \propto \tilde{J}$ and $\Delta E_B \equiv { \sum_{k}}\omega_{m_k}-\omega_{n_k}={ \sum_{k}}(m_k-n_k) \omega_{k}$. If we make an assumption that $\frac{\Delta E^f}{\Delta E_B} <<1$\cite{29,54,55,56,57}, which implies $\frac{\tilde{J}} {\Delta E_B}<<1$, we can ignore   the $E^f_{ p}-E^f_j$ term in  exponential of the above equation \ref{LFH}. 
Thus in this situation, known as anti-adiabatic approximation, we can safely ignore the time dependence of the system operators. This approximation is the reminiscent of Markovian approximation. This equation therefore simplifies to 

\begin{eqnarray}
	\label{ad}
	{	H}^{\prime}_I(t)&=&  \sum_{\{m_k\},\{n_k\}} e^{-i (\omega_{m_k}-\omega_{n_k})t}  |m_k\rangle \langle n_k| \langle m_k | {H}^{\prime}_I  |n_k\rangle = \tilde{J}[ a^{\dagger}_1 a_2\mathcal{B}^{\dagger}(t)+ a^{\dagger}_2 a_1 \mathcal{B}(t)],
\end{eqnarray}
where $\mathcal{B}^\dagger(t)= e^{i{H}^{\prime}_B t}\mathcal{B}e^{-i{H}^{\prime}_Bt} $ is the time evolved modified bath operator. Using this form of the time evolved interaction Hamiltonian, we get the following master equation for the system density operator { at $ 0K$ temperature}(See Appendix B for details):

\begin{eqnarray}
	\label{ME}
	\dot{\rho}_S(t)&=& \gamma_+(t)[ L^{\dagger}_{12} \rho_S L_{12}- \frac{1}{2}\{ L_{12}L_{12}^{\dagger}, \rho_S \}] + \gamma_+(t)[ L_{12} \rho_S L_{12}^{\dagger}- \frac{1}{2}\{ L_{12}^{\dagger}L_{12}, \rho_S \}] 
	\nonumber \\
	&&+\gamma_-(t) [L_{12} \rho_S L_{12} L_{12}^{\dagger}\rho_S L_{12}^{\dagger}],
\end{eqnarray}
where $L_{ij}=a^{\dagger}_i a_j$ are system operators governing the dynamics, and  the decoherence functions $\gamma_{\pm}(t)$ are given by
\begin{eqnarray}
&&	\gamma_{\pm}(t)= 2\tilde{J}^2 \int_0^t d\tau \Bigg[ e^{\pm \sum_k |g_{1k}-g_{2k}|^2 \cos [\omega_k (t-\tau)] } \cos \Bigg(\sum_k |g_{1k}-g_{2k}|^2 \sin [\omega_k (t-\tau)]\Bigg) -1 \Bigg].
	\end{eqnarray}
Next, we solve this master equation \ref{ME} for a given initial density matrix in the single particle basis $\{ a_1^{\dagger}|0\rangle, a_2^{\dagger}|0\rangle\}$. 
	 Let $|S\rangle=\frac{1}{\sqrt{2}}[|10  \rangle - |01\rangle] $ be the singlet state and the triplet state $|T\rangle= \frac{1}{\sqrt{2}}[|10 \rangle + |01\rangle] $, $|0\rangle$ means the empty site while $|1\rangle$ means site filled with a single particle. In spin language, $|S\rangle$ and $|T\rangle $ translate to the states with $|S^T=0,S^Z=0\rangle$ and $|S^T=1,S^Z=0\rangle$   respectively. $S^T$ is the total spin angular momentum with $S^z$ to be its  $z$-component. In this singlet-triplet basis, the diagonal and off-diagonal terms evolve as:
	
	
	\begin{figure}[t]
		\includegraphics[width=5.5cm,height=5cm]{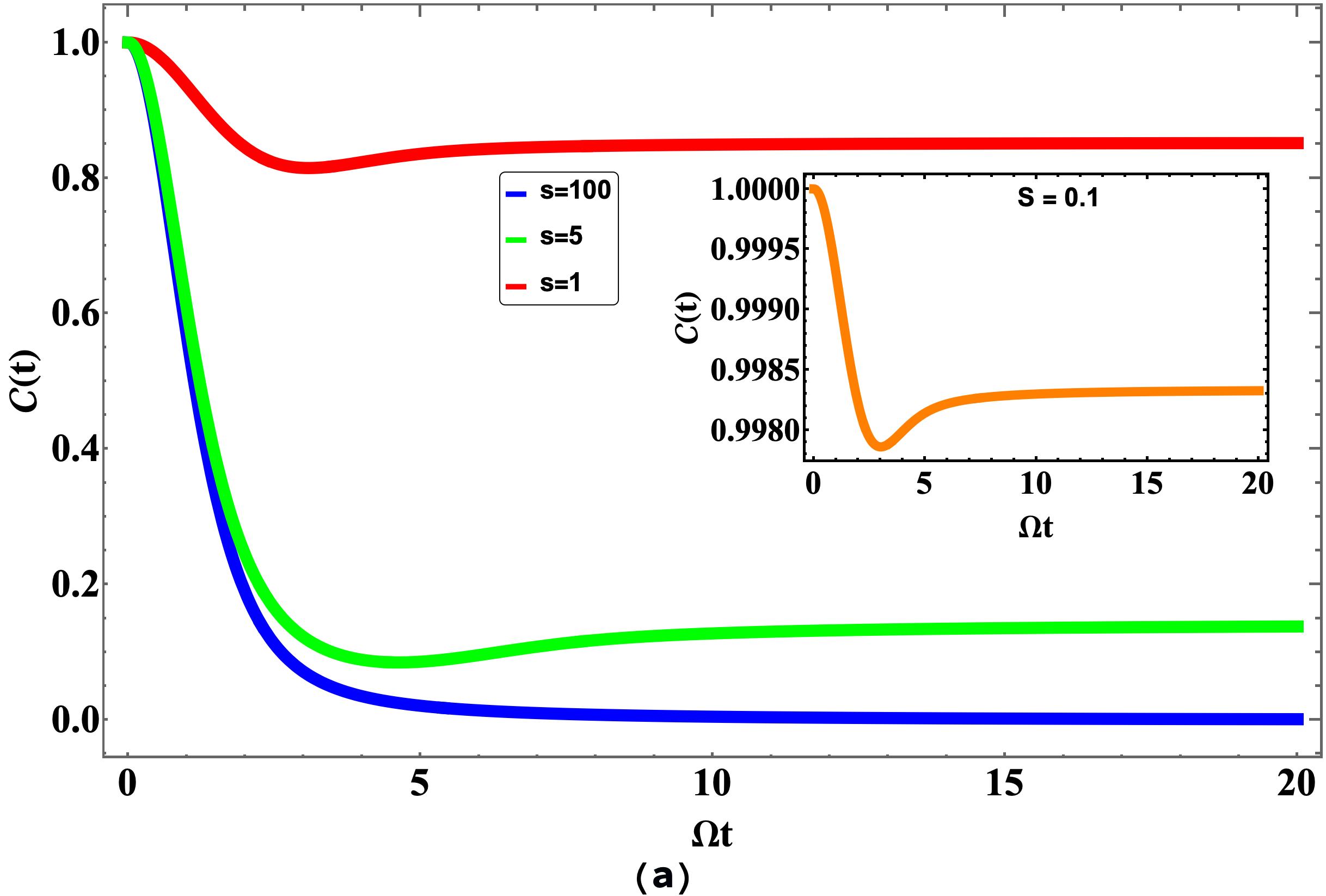} \hspace{0.31cm}
		\includegraphics[width=5.5cm,height=5cm]{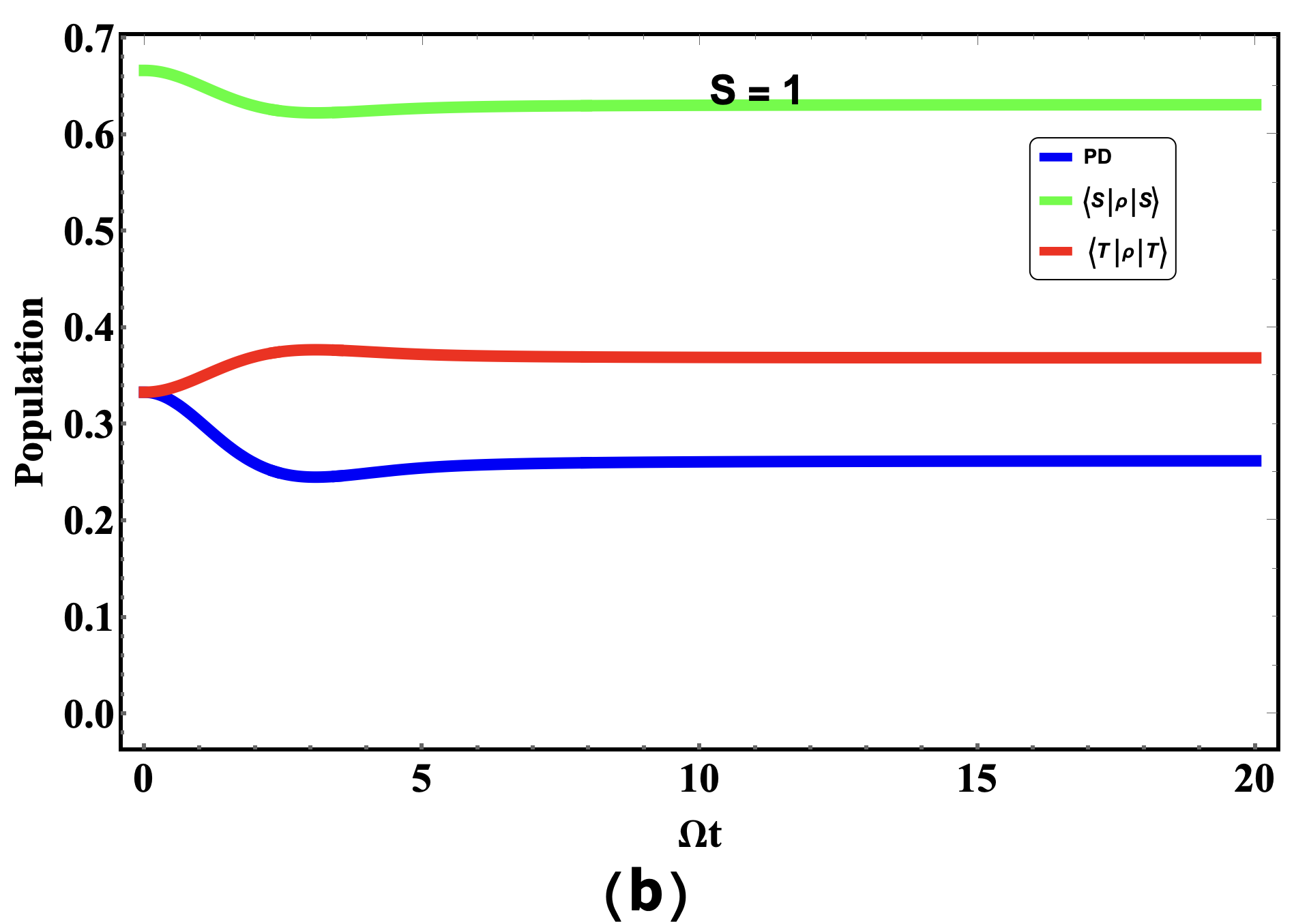} \\
		\includegraphics[width=5.5cm,height=5cm]{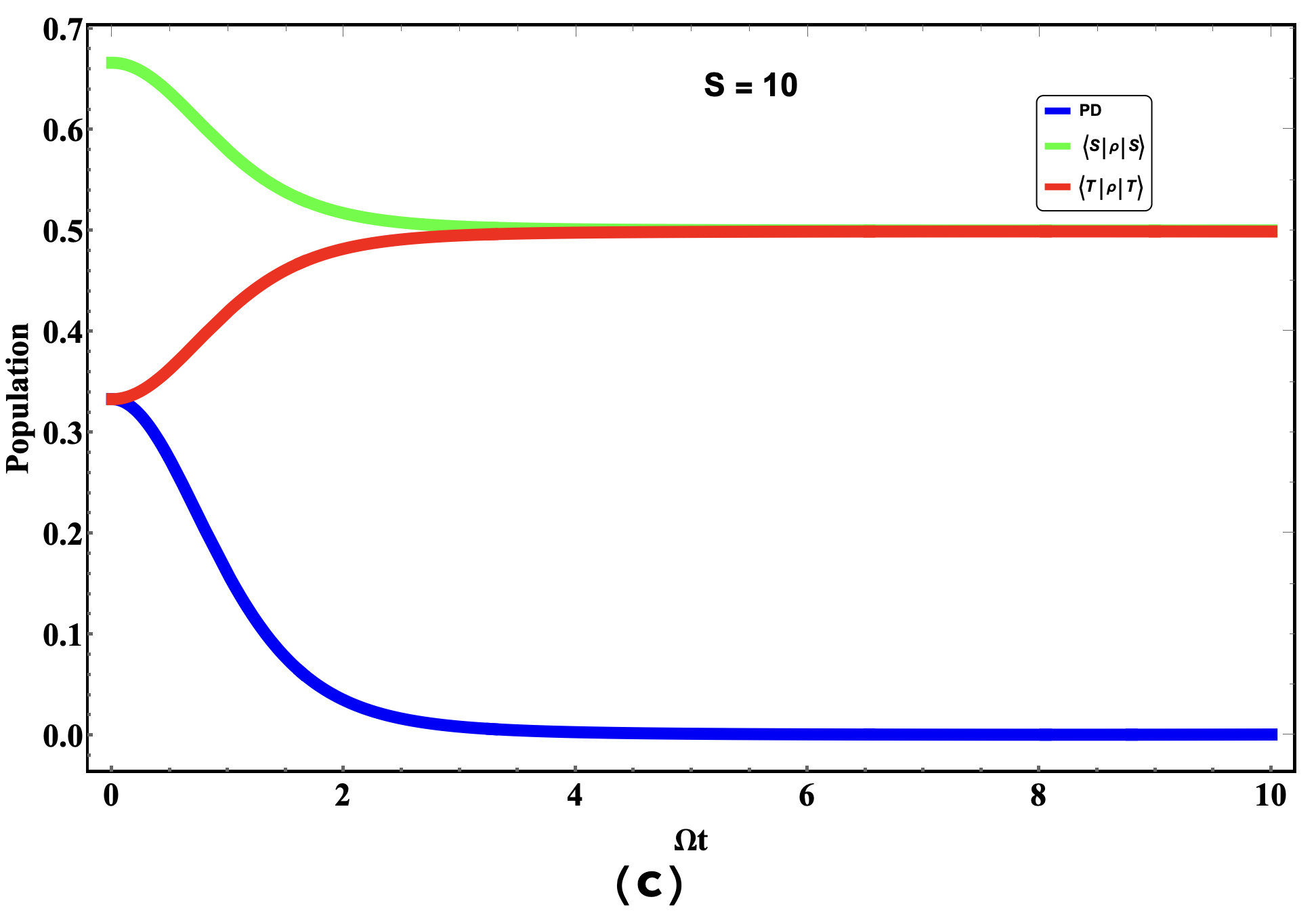} \hspace{0.31cm}
		\includegraphics[width=5.5cm,height=5cm]{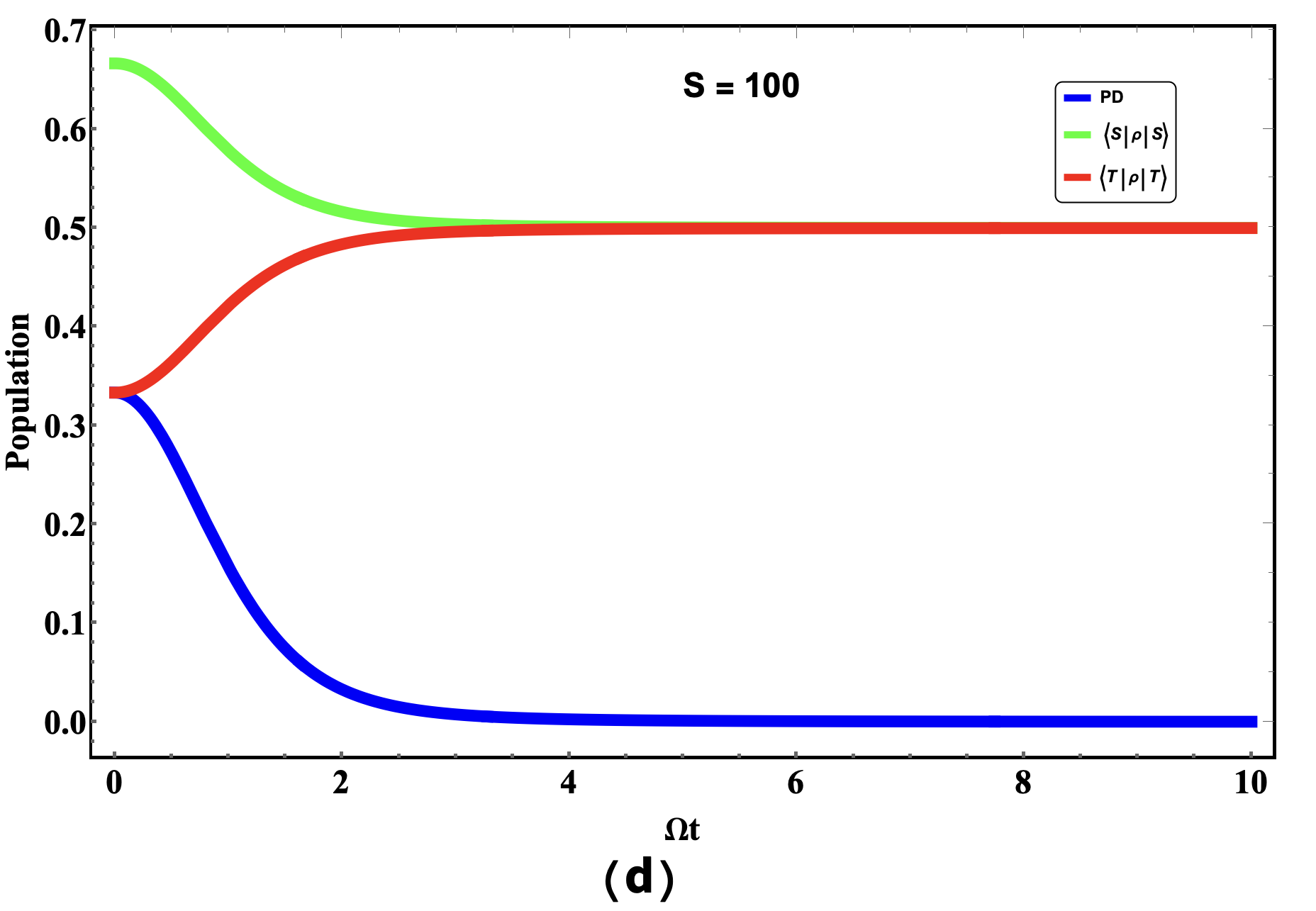} 
		\caption{Here (a) represents the decay of off-diagonal elements of the density matrix  quantified by $\mathcal{C}$(t) for different values of $s$ for the initial state$|\psi\rangle=\sqrt{\frac{2}{3}}|S\rangle + \sqrt{\frac{1}{3}}|T\rangle$. (b)-(d) represent the variation of population difference $\mathcal{P}_D(t)$ and  the populations of the $|T\rangle$ and $|S\rangle$ states for the same initial state $|\psi\rangle$.}
		\label{coh}
	\end{figure}
	\begin{eqnarray}
			\label{DFE}
		\frac{d}{dt}\langle T|{\rho_s(t)}|T\rangle &= & -(\frac{2\gamma_+(t) -\gamma_-(t)}{2}) [\langle T|{\rho_s}(t)|T \rangle-\langle S|{\rho_s}(t)| S \rangle]  \\  	
		\label{DF}
		\frac{d}{dt}\langle S|{\rho_s(t)}|S \rangle &= &-(\frac{2\gamma_+(t) -\gamma_-(t)}{2})[\langle S|{\rho_s}(t)|S \rangle-\langle T|{\rho_s}(t)|T\rangle]
		\\
		\label{D}
		\frac{d}{dt}\langle T|{\rho_s(t)}|S\rangle &= &-(\frac{\gamma_-(t)+6\gamma_+(t)}{2})\langle T|{\rho_s}(t)|S\rangle - (\frac{\gamma_-(t)-2\gamma_+(t)}{2})\langle S|{\rho_s}(t)|T\rangle.
	\end{eqnarray}
	The equations \ref{DFE} ,\ref{DF} ,\ref{D} can be solved to yield the following dynamics of diagonal and off-diagonal terms:		
	\begin{eqnarray}
		\langle S|\rho_s(t)|S\rangle &=& \frac{1}{2}\langle S|\rho_s(0)|S\rangle[1+\exp[-\int_0^t ds \Gamma_0(s)]] +  \frac{1}{2}\langle T|\rho_s(0)|T\rangle[1-\exp[-\int_0^t ds \Gamma_0(s)]] \\
		\langle S|\rho_s(t)|T\rangle &=& \frac{1}{2}\langle S|\rho_s(0)|T\rangle [e^{- \int_0^t ds \Gamma_1(s)}+ e^{-\int_0^t ds \Gamma_2(s)} ] +  \frac{1}{2}\langle T|\rho_s(0)|S\rangle [e^{- \int_0^t ds \Gamma_1(s)}-e^{-\int_0^t ds \Gamma_2(s)} ],
	\end{eqnarray}
	where $ \Gamma_0(t)=\frac{2\gamma_+(t) -\gamma_-(t)}{2} $, $\Gamma_1(t)= 2\gamma_+(t)+ \gamma_-(t)$ and  $ \Gamma_2(t)= 4\gamma_+(t)$. 
	Next, we simplify these solutions using the  Ohmic bath spectral density with Gaussian cut-off as introduced above that has the form $|g(\omega)|^2=\alpha \omega e^{\frac{-\omega^2}{\Omega^2}}$. Further, we introduced   $\vec{l}=\vec{r}_1-\vec{r}_2$ to  be the distance of separation between  two sites with $\vec{r}_1$ and $\vec{r}_2$ to be their respective position vectors. Therefore, we write 
	\begin{eqnarray}
		\label{SPD}
		\sum_{k}|\alpha_k|^2\cos{\omega_k(t-\tau)} &=& 2\sum_{k}\frac{|g_k|^2}{\omega^2_{k}}[1-\cos{(\vec{k}.\vec{l})}]\cos\omega_{k}(t-\tau)) \nonumber \\
		&=& \frac{\alpha}{v^3 \pi^2}\int_{0}^{\infty}  d\omega \omega e^{\frac{-\omega^2}{\Omega^2}} (1-\frac{\sin \omega s}{\omega s})\cos\omega(t-\tau).
	\end{eqnarray}
	In the second line, we have integrated over solid angle $\theta$ and $\phi$, and used the linear dispersion $\omega=v k$, where $v$ is the speed of phonons. $s=\frac{l}{v}$ defines an intrinsic scattering time scale. Thus, we can tune the values of $s$ by changing the separation $l$ between the lattice sites{ \cite{58,59}}.  From these equations \ref{SPD}, we observe that, if $\vec{k}.\vec{l}=2n\pi$, $n=0,1,2,...,$ we have $\gamma_{\pm}(t)=0$, thus showing no decoherence in the system. This implies there exist certain bath modes in the forward scattering process that do not couple to  system thus enabling the qubit  to maintain coherence. { We  have the factor in equation \ref{SPD} as $1-\cos{(\vec{k}.\vec{l})}=\sin^2(\vec{k}.\vec{l})$. This factor suppresses the effect of bath modes on the qubit for certain values of $(\vec{k}.\vec{l})$. It means there exist certain bath modes which do not resolve the separation of the lattice sites and hence causes less coherence.}  Since, there exist different energy scales in our model, we assume energy cutoff $\Omega$ to provide highest energy scale and all variables are measured with respect to it: $\omega \rightarrow\frac{\omega}{\Omega},~t \rightarrow \Omega t,~ s\rightarrow \Omega s$. Without loss of generality, we put $\Omega=1$. Also, from the  experimental point of view, for example in cold atom settings, the distance of separation of the two sites can be varied from $l=100 \times 10^{-9}$m to $10 \times 10^{-6}$m with the speed of phonons to be around $ v=350 m s^{-1}$, thus tuning scattering scale from $s=10^{-10}$s to $10^{-8}$s. 

	Next, in order to analyze the decoherence in our system, we define  normalized coherence by 
	$\mathcal{C}(t)=\frac{|\langle T|\rho(t)|S\rangle|}{|\langle T|\rho(0)|S\rangle|}$, that represents the loss of off-diagonal terms in the density matrix. Also, the population difference of singlet and triplet states  is given by  $\mathcal{P}_D(t)= \frac{|\langle T|\rho(t)|T\rangle-\langle S|\rho(t)|S\rangle|}{|\langle T|\rho(0)|T\rangle+\langle S|\rho(0)|S\rangle|}$ .
In order to look at the behaviour of $\mathcal{C}(t)$ and $\mathcal{P}_D(t)$ for different values of $s$, without loss of generality, we choose an initial state of the form $|\psi\rangle=\sqrt{\frac{2}{3}}|S\rangle + \sqrt{\frac{1}{3}}|T\rangle$. 	
	 We plot time evolution of $\mathcal{C}(t)$ and $\mathcal{P}_D(t)$  in figure \ref{coh}. From figure \ref{coh}(a), we observe that $\mathcal{C}(t)$ behaves differently for different values of $s$. Thus for small values of $s$, the coherence $\mathcal{C}(t)$ is maintained for longer times while for large values of $s$, i.e. $s>>1$, the system loses coherence and state of the qubit becomes completely  decoherent in the long time limit. This behaviour can be attributed to the localization-delocalization  effects in the model and graphically represented in figure \ref{coh}(b)-(d). Here, we have  time variation of $\mathcal{P}_D(t)$ for the same initial state. We observe from these plots that the  initial delocalized state $|\psi\rangle$  gets localized over certain time period as we increase value of $s$. For small values of $s=1$ (fig. \ref{coh}(b)), the system remains mostly in the energy eigen states $|S\rangle$ and $|T\rangle$, thus making system more coherent. While, as we increase $s=10, ~100$ in figures \ref{coh}(c)-(d), the system goes over to the probabilistic mixture of $|S\rangle$ and $|T\rangle$ with equal probabilities. This means the fermion gets localized into one of the lattices sites, thus yielding a decoherent state.
	 In nutshell, as we decrease the distance between two sites, the available phase space volume for the scattering of phonons will get reduced making less number of phonons to interact causing small decoherence in the system. The available phase space  volume increases with increase in $s$, causing the particle to get localized into of the sites, therefore we have equal probable mixture of $|S\rangle$ and $|T\rangle$ states.

	\section{ Decoherence Control with $\pi$-pulses}
	
		In this section, we consider a protocol of controlling decoherence { (in original frame of refrence)} known as bang-bang control{ \cite{45,46,49}}. In this protocol, the system is decoupled from its bath using the sequences of suitably tailored pulses through unitary operations. We consider a pulse that flips the states at two sites simultaneously, the corresponding operator is  
	defined by $\Pi=a_1^\dagger a_2 + a_2^\dagger a_1 $. Here we restrict our analysis to the single particle subspace so that the following property is satisfied: $\Pi^2 |\phi\rangle=|\phi\rangle$, where $|\phi\rangle \epsilon \{|S\rangle, |T\rangle \}$. 
	These pulses are applied over an infinitesimal time $\delta t$, so that the time evolution for one complete cycle is given by
	\begin{eqnarray}
		\mathcal{U}_{cycle}(t,\delta t)= \mathcal{U}_I(t+2 \delta, t+\delta {  t}) \Pi \mathcal{U}_I(t+\delta t,t) \Pi,
	\end{eqnarray}
	where $\mathcal{U}_I(t,t')$ is the interaction picture time evolution operator and can be written in the following way:
	\begin{eqnarray}
		\mathcal{U}_{ I}(t,t')&=&\mathcal{U}_I(t,0)  \mathcal{U}_I^\dagger(t',0) \nonumber	\\
		&=&e^{iH_0 t} e^{{ -}iH t}e^{iH t'}e^{-iH_0 t'}.
	\end{eqnarray}
Using the fact that $[H_0, \Pi]=0$, therefore, we write the time evolution operator for the complete cycle as
	\begin{eqnarray}
		\label{PP}
		\mathcal{U}_{cycle}(t,\delta t)&=& \mathcal{U}(t+2\delta t,t+\delta t)\Pi \mathcal{U}(t+\delta t,t)\Pi
	\nonumber	\\&=&e^{iH_0(t+2\delta t)} e^{-iH\delta t}e^{-iH_0(t+\delta t)}\Pi e^{iH_0(t+\delta t)} e^{-iH\delta t}\Pi e^{-iH_0 {  t}} \Pi
		\nonumber\\&=&e^{iH_0(t+2\delta t)}e^{-iH\delta t}\Pi e^{-iH\delta t}\Pi e^{-iH_0t}{ \Pi}.
	\end{eqnarray}
We further simplify the above equation \ref{PP}, using the fact $ \Pi (H_0+H_I) \Pi =H_0+{ \tilde{H}_I} $, where $ { \tilde{H}_I}= n_1 B_2 + n_2 B_1$, $ B_{ p}=\sum_k  (g_{{ p}k} b_k + g^{\star}_{{ p}k} b^{\dagger}_k) $. Using Baker-Campbell-Hausdorff (BCH) formula $e^X e^Y =e^{X+Y+\frac{1}{2} [X,Y]+...}$ we have up to $O(\delta t^2)$

	\begin{eqnarray*}
		\mathcal{U}_{cycle}(t,\delta t)&=&e^{iH_0(t+2\delta t)} e^{-i(H_0+H_I)\delta t} e^{-i(H_0+{ \tilde{H}_I})\delta t} e^{-iH_0 t} \\&=&e^{iH_0(t+2\delta t)} e^{-i(2H_0+H_I+{ \tilde{H}_I})\delta t -\frac{1}{2}[H,H_0+{ \tilde{H}_I}]\delta t^2} e^{-iH_0 t}
		\\&=& e^{iH_0(t+2\delta t)} e^{-i(2H_0+B_1 +B_2)\delta t -\frac{1}{2}[H,H_0+{ \tilde{H}_I}]\delta t^2} e^{-iH_0 t}
	\end{eqnarray*}
	Again, using $e^{X+Y}=e^X e^Y e^{\frac{1}{2}[X,Y]+...}$ upto $O(\delta t ^2)$ , we write
	\begin{eqnarray}
		\mathcal{U}_{cycle}(t,\delta t)&{ \approx}& e^{iH_0(t+2\delta t)} e^{-i(2H_0+B_1 +B_2)\delta t} e^{-\frac{1}{2}[H,H_0+{ \tilde{H}_I}]\delta t^2} e^{-iH_0 t}\nonumber \\&{ \approx}&e^{iH_0(t+2\delta t)} e^{-2i H_0\delta t}  e^{-i(B_1 +B_2)\delta t} e^{\frac{1}{2} [2H_0, B_1+B_2]\delta t ^2}e^{-\frac{1}{2}[H,H_0+{ \tilde{H}_I}]\delta t^2}e^{-iH_0 t}\nonumber\\&{ \approx}& e^{iH_0 t} e^{-i(B_1 +B_2)\delta t}\{1+\delta t ^2[H_B, B_1 +B_2]\}\{1-\frac{1}{2} \delta t^2[H,H_0+{ \tilde{H}_I}]\}^{-iH_0 t}
	\nonumber	\\&{ \approx}& e^{iH_0 t} e^{-i(B_1 +B_2)\delta t}\{1-\frac{1}{2} \delta t^2[H,H_0+{ \tilde{H}_I}]+\delta t ^2[H_B, B_1 +B_2]\}e^{-iH_0 t}+O (\delta t ^4)\nonumber\\&{ \approx}&e^{iH_0 t} e^{-i(B_1 +B_2)\delta t}\{1- \delta t ^2 \hat{C} \}e^{-iH_0 t},
	\end{eqnarray}
	Where, $\hat{C}=\frac{1}{2}[H,H_0+{ \tilde{H}_I}]{ -}[H_B, B_1 +B_2]$. The time evolved  density matrix of the  system is
	\begin{eqnarray}
		\label{SDM}
		{\rho}_S( \delta t)&=&\sum_{\{n_k\}} \langle \{n_k\}|	\mathcal{U}_{cycle}(t,\delta t)\big({\rho}_S(0)\otimes|0 \rangle \langle0|\big)	\mathcal{U}^\dagger_{cycle}(t,\delta t)|\{n_k\}\rangle \nonumber  \\
		&{ \approx}& \sum_{\{n_k\}} \langle \{n_k\}| 	e^{iH_0 t} e^{-i(B_1 +B_2)\delta t}\{1-\hat{C} \delta t ^2\}e^{-iH_0 t} \big({\rho}_S(0)\otimes|0 \rangle \langle0|\big) e^{iH_0 t}  \{1-\hat{C}^{\dagger} \delta t ^2\}e^{i(B_1 +B_2)\delta t}   e^{-iH_0 t}     |\{n_k\}\rangle \nonumber \\
		&{ \approx}& \sum_{\{n_k\}} \langle \{n_k\}| e^{iH_0 t} e^{-i(B_1 +B_2)\delta t} e^{-iH_0 t} |0\rangle \rho_S(0) \langle 0| e^{iH_0 t} e^{i( B_1 +B_2)\delta t}   e^{-iH_0 t}     |\{n_k\}\rangle \nonumber \\
		&& -\delta t^2 \Bigg[ \sum_{\{n_k\}} \langle \{n_k\}| e^{iH_0 t} e^{-i(B_1 +B_2)\delta t} \hat{C}e^{-iH_0 t} |0\rangle \rho_S(0)\langle 0| e^{iH_0 t} e^{i(B_1 +B_2)\delta t}   e^{-iH_0 t}     |\{n_k\}\rangle + H.C. \Bigg] + O(\delta t^4).
	\end{eqnarray}
Next, we simplify these expressions. The first term in the above equation \ref{SDM} can be simplified as
\begin{eqnarray}
	&&\sum_{\{n_k\}} \langle \{n_k\}|	e^{iH_0 t} e^{-i(B_1 +B_2)\delta t} e^{-iH_0 t}|0\rangle \rho_S(0) \langle 0|	e^{iH_0 t} e^{i(B_1 +B_2)\delta t} e^{-iH_0 t}|\{n_k\}\rangle \nonumber \\
	&&~~~~~~~~~~~~~~~~~~~~~~~~~~~ = \sum_{\{n_k\}}  \langle \{n_k\}|	e^{iH_B t}  e^{-i(B_1 +B_2)\delta t}|0\rangle \rho_S(0) \langle 0| e^{i(B_1 +B_2)\delta t} e^{-iH_B t}|\{n_k\}\rangle \nonumber\\
	&&~~~~~~~~~~~~~~~~~~~~~~~~~~~~= \langle 0|e^{-i(B_1 +B_2)\delta t}e^{i(B_1 +B_2)\delta t}|0\rangle \rho_S(0) = \rho_S(0).
\end{eqnarray}
Similarly, second term in equation (\ref{SDM})yields
\begin{eqnarray}
	&&\sum_{\{n_k\}} \langle \{n_k\}|	e^{iH_0 t} e^{-i(B_1 +B_2)\delta t} e^{-iH_0 t}|0\rangle \rho_S(0) \langle 0|	e^{iH_0 t} \hat{C}^\dagger e^{i(B_1 +B_2)\delta t} e^{-iH_0 t}|\{n_k\}\rangle \nonumber 	\\
	&&~~~~~~~~~~~~= \sum_{\{n_k\}}  \langle \{n_k\}|	e^{iH_B t}  e^{-i(B_1 +B_2)\delta t}   \rho_S(0) \otimes|0\rangle \langle 0| e^{iH_0 t} \hat{C}^\dagger e^{i(B_1 +B_2)\delta t} e^{-iH_0 t}|\{n_k\}\rangle \nonumber \\
	&&~~~~~~~~~~~~=Tr_B \bigg[e^{iH_B t}  e^{-i(B_1 +B_2)\delta t}   \rho_S(0) \otimes|0\rangle \langle 0| e^{iH_0 t} \hat{C}^\dagger e^{i(B_1 +B_2)\delta t} e^{-iH_0 t}\bigg] \nonumber   \\
	&&~~~~~~~~~~~~=Tr_B \bigg[  ( \rho_S(0) \otimes|0\rangle \langle 0| )e^{iH_0 t} \hat{C}^\dagger e^{i(B_1 +B_2)\delta t} e^{-iH_S t} e^{-i(B_1 +B_2)\delta t}\bigg] \nonumber \\ 
	&&~~~~~~~~~~~~=Tr_B \bigg[  ( \rho_S(0) \otimes|0\rangle \langle 0| )e^{iH_0 t} \hat{C}^\dagger e^{-iH_S t} \bigg] \nonumber \\
	&&~~~~~~~~~~~~= \rho_S(0) \otimes \langle 0| e^{iH_0 t} \hat{C}^\dagger e^{-iH_S t}|0\rangle  = \rho_S(0) \otimes e^{iH_S t}\langle 0|  \hat{C}^\dagger|0\rangle e^{-iH_S t} = 0.
\end{eqnarray}
Using these results into equation \ref{SDM}, we have  up to $O(\delta t^4)$, $\rho_S(\delta t)= \rho_S(0)$. Next, we use  the relation $\delta t N=T$ and repeat the cycle  $N$-times i.e. for finite duration $T$,  we write the evolution of the density matrix 
\begin{eqnarray}
	\rho_S(T)= \underset{N-factors}{\underbrace{\rho_S(\delta t).....\rho_S(\delta t)}}= \rho_S(0).
\end{eqnarray}
Thus, applying simultaneously tailored $\pi$ -pulses, we are able to control the decoherence in the system. { Therefore up to $O(\delta t^3)$, we see that system does not undergo decoherence in the finite interval of time.}

	\section{Conclusions}
	In conclusion, we studied a model of a qubit constructed from a spinless fermion hopping between two lattices sites, while these lattice sites are strongly coupled to a collective dephasing bath. To work perturbatively, we transformed the total system via Lang-Firsov transformation to a polaron frame, where the system-bath coupling gets substantially reduced. In this dressed basis, we solved for the diagonal and off-diagonal terms of the system density matrix in the singlet-triplet basis. We identified an intrinsic time scale $s$ (or a length scale) that helps to manipulate the decoherence in our model. The large values of $s$, the system loses coherence completely while for the small values of $s$, the system maintains the coherence for longer times. This is also reflected in the  dynamics of probabilities of the singlet and triplet states. For small values of $s$, if the system starts in the one of the states, it stays in the given state while for large value of $s$, it saturates into the equal probable mixture of singlet and triplet states. Thus tuning $s$, we can set the system into localization-delocalization transition. 

Furthermore, we present a way to prevent decoherence and exercise quantum control by simultaneously creating and annihilating the particle at a lattice site with an externally administered fast train of pulses.  Notably, we have observed that the system remains free from decoherence with error upto $\delta t^{ 3}$ for $N$ steps during the described evolution. By selecting arbitrarily high values of $N$, we can achieve an arbitrarily small error in the preservation of quantum information.

 \begin{acknowledgements}
SB would like to  thank   DST Govt. of India for financial assistance through INSPIRE fellowship no. DST/INSPIRE Fellowship/[IF210401].
\end{acknowledgements}

	{\appendix
	\section{Lang-Firsov Transformation}
	In this appendix, we provide detailed calculation of obtaining transformed Hamiltonian using Lang-Firsov (LF) transformation.
	In the transformed frame, we write $H^{\prime}= e^S(H_S+H_B+H_I)e^{-S}$ with $S=-\sum_{{ p},k}n_{ p}\bigg(\frac{g_{{ p}k} }{\omega_k}b_k-\frac{g_{{ p}k}^*}{\omega_k }b_k^\dagger \bigg)$. Therefore, for system operators $a_i$, with the help of Baker-Campbell-Hausdorff (BCH) formula, we write
	
	\begin{eqnarray}
		e^Sa_{ p}^{\dagger} e^{-S} &=&e^{-\sum_{j,k}n_j\bigg(\frac{g_{jk} }{\omega_k}b_k-\frac{g_{jk}^{\star}}{\omega_k }b_k^\dagger \bigg)} a_{ p}^{\dagger} e^{\sum_{j,k}n_j\bigg(\frac{g_{jk} }{\omega_k}b_k-\frac{g_{jk}^{\star}}{\omega_k }b_k^\dagger \bigg)} \nonumber \\
		&=& 	
		a_{ p}^{\dagger}-\Big[\sum_{j,k}n_j\bigg(\frac{g_{jk} }{\omega_k}b_k-\frac{g_{jk}^{\star}}{\omega_k }b_k^\dagger \bigg),a_{ p}^{\dagger}\Big]
		+\frac{1}{2!}\bigg[\sum_{j,k}n_j\bigg(\frac{g_{jk} }{\omega_k}b_k-\frac{g_{jk}^{\star}}{\omega_k }b_k^\dagger \bigg),\Big[\sum_{l,m}n_l\bigg(\frac{g_{lm} }{\omega_m}b_k-\frac{g_{lm}^{\star}}{\omega_m }b_m^\dagger \bigg),a_{ p}^\dagger\Big]\bigg]+....\nonumber \\
		&=& a_{ p}^\dagger-\sum_{k}\bigg(\frac{g_{{ p}k} }{\omega_k}b_k-\frac{g_{{ p}k}^{\star}}{\omega_k }b_k^\dagger\bigg)a_{ p}^\dagger+ \frac{1}{2!}\bigg(\sum_{k}\frac{g_{{ p}k} }{\omega_k}b_k-\frac{g_{{ p}k}^{\star}}{\omega_k }b_k^\dagger\bigg)^2 a_{ p}^\dagger+... \nonumber \\
		&=& e^{-\sum_{k}\bigg(\frac{g_{{ p}k} }{\omega_k}b_k-\frac{g_{{ p}k}^{\star}}{\omega_k }b_k^\dagger\bigg)}a_{ p}^\dagger,
	\end{eqnarray}
		where we have used $[n_{ p}, a_j^\dagger]=a_j ^\dagger\delta_{{ p}j}$\\
	Therefore,
	\begin{eqnarray}
		\label{SP}
		e^S H_S  e^{-S}=J \bigg[e^{-\sum_{k}\bigg(\frac{g_{1k} }{\omega_k}b_k-\frac{g_{1k}^{\star}}{\omega_k }b_k^\dagger\bigg)}e^{\sum_{k}\bigg(\frac{g_{2k} }{\omega_k}b_k-\frac{g_{2k}^*}{\omega_k }b_k^\dagger\bigg)}a_1^\dagger a_2+e^{\sum_{k}\bigg(\frac{g_{1k} }{\omega_k}b_k-\frac{g_{1k}^{\star}}{\omega_k }b_k^\dagger\bigg)}e^{-\sum_{k}\bigg(\frac{g_{2k} }{\omega_k}b_k-\frac{g_{2k}^{\star}}{\omega_k }b_k^\dagger\bigg)}a_2^\dagger a_1 \bigg]
	\end{eqnarray}
Also,
	
	\begin{eqnarray*}
		e^S b^\dagger_k  e^{-S}
		&=& e^{-\sum_{{ p},k'} n_{ p} (\frac{g_{{ p}{k'}}}{\omega_{k'}}b_{k' }- \frac{g^{\star}_{{ p}k'}}{\omega_{k'}}b^\dagger_{k'})}  b^\dagger_k e^{\sum_{{ p},k'} n_i (\frac{g_{{ p}{k'}}}{\omega_{k'}}b_{k' }- \frac{g^*_{{ p}{k'}}}{\omega_{k'}}b^\dagger_{k'})}\\
		&=& b_{k}^\dagger-\bigg[\sum_{i,k'} n_{ p} (\frac{g_{{ p}k'}}{\omega_{k'}}b_{k' }- \frac{g^{\star}_{{ p}k'}}{\omega_{k'}}b^\dagger_{k'}),b_k^{\dagger}\bigg]+..... \\
		&=&b_k^\dagger-\sum_{{ p},k}n_{ p} \frac{g_{{ p}k}}{\omega_k}
	\end{eqnarray*}		
	Therefore,
	\begin{eqnarray} 
		e^SH_Be^{-S}&=&\sum_k \omega_k \bigg[b_k^\dagger-\sum_{i}n_{ p} \frac{g_{{ p}k}}{\omega_k}\bigg]\bigg[b_k-\sum_{j}n_j \frac{g_{jk}^{\star}}{\omega_k}\bigg]\nonumber\\
		&=&\sum_{k} \omega_k  b_k^\dagger b_k-\sum_{j,k}n_jg_{jk}^{\star} b_k^\dagger-\sum_{{ p},k}n_{ p} g_{{ p}k} b_k+\sum_{{ p},j,k}n_{ p} n_j \frac{g_{{ p}k} g_{jk}^{\star}}{\omega_k}
	\end{eqnarray}
	
	In a similar fashion, the transformed interaction Hamiltonian becomes
	\begin{eqnarray}
		e^S H_I e^{-S}&=& e^{-\sum_{l,k} n_l(\frac{g_{lk}}{\omega_k}b_k - \frac{g^{\star}_{lk}}{\omega_k}b^\dagger_k )}\sum_{{ p},k} n_{ p}  (g_{{ p}k}b_k + g^{\star}_{{ p}k} b^\dagger_k)e^{\sum_{l,k} n_l (\frac{g_{lk}}{\omega_k}b_k - \frac{g^{\star}_{lk}}{\omega_k}b^\dagger_k )}\nonumber	\\
		&=&\sum_{{ p},k}n_{ p}( g_{{ p}k}b_k+ g_{{ p}k}^{\star} b_k^\dagger )-2\sum_{{ p},k,l}n_{ p} n_l \frac{g_{{ p}k}^{\star}g_{lk}}{\omega_k}
	\end{eqnarray}
	Therefore, we write,
	\begin{eqnarray}
		\label{TE}
		e^S [H_B+H_I]e^{-S}=\sum_{k} \omega_k b_k^\dagger b_k-\sum_{{ p},k,j}n_{ p} n_j \frac{g_{{ p}k}^{\star}g_{jk}}{\omega_k}.
	\end{eqnarray}
	
	Lets calculate the second term of above equation \ref{TE} we have,
	\begin{eqnarray*}
		\sum_{{ p},k,j}n_{ p} n_j \frac{g_{{ p}k}^{\star}g_{jk}}{\omega_k}&=&\sum_{{ p},k,j}n_{ p} n_j \frac{g_{{ p}k}^{\star}g_{jk}+g_{ik}g_{jk}^{\star}}{2 \omega_k}
		\\&=&\sum_{k}n_1n_1 \frac{g_{1k}^{\star}g_{1k}+g_{1k}g_{1k}^{\star}}{2 \omega_k}+2\sum_{k}n_1n_2 \frac{g_{1k}^{\star}g_{2k}+g_{1k}g_{2k}^{\star}}{2 \omega_k}+\sum_{k}n_2 n_2 \frac{g_{2k}^{\star}g_{2k}+g_{2k}g_{2k}^{\star}}{2 \omega_k}
		\\&=&=\sum_{k}n_1^2 \frac{|g_{1k}|^2}{\omega_k}+\sum_{k}n_1n_2 \frac{g_{1k}^{\star}g_{2k}+g_{1k}g_{2k}^{\star}}{ \omega_k}+\sum_{k}n_2^2 \frac{|g_{2k}|^2}{\omega_k}\\
		&=&\sum_{k}n_1^2 \frac{|g_{1k}|^2}{\omega_k}+\sum_{k}n_2^2 \frac{|g_{2k}|^2}{\omega_k}+V_{12}n_1 n_2
	\end{eqnarray*}
	where,  $V_{12}=\sum_{k} \frac{g_{1k}^*g_{2k}+g_{1k}g_{2k}^*}{ \omega_k}$.
\\
	
	Therefore, equation \ref{TE} becomes
	\begin{eqnarray}
		e^S [H_B+H_I]e^{-S}= \sum_{k} \omega_k b_k^\dagger b_k-\bigg[\sum_{k} (n_1^2\frac{|g_{1k}|^2}{\omega_k}+ n_2^2\frac{|g_{2k}|^2} {\omega_k})+V_{12}n_1 n_2\bigg].
	\end{eqnarray}
	
	Therefore, by adding equation \ref{SP} and \ref{TE}, the total transformed Hamiltonian $H'$ becomes\\
	\begin{eqnarray}
		\label{TEQ}
		H'&=&e^S [H_S+H_B+H_I]e^{-S}\nonumber
		\\&=&J \bigg[e^{-\sum_{k}\bigg(\frac{g_{1k} }{\omega_k}b_k-\frac{g_{1k}^{\star}}{\omega_k }b_k^\dagger\bigg)}e^{\sum_{k}\bigg(\frac{g_{2k} }{\omega_k}b_k-\frac{g_{2k}^{\star}}{\omega_k }b_k^\dagger\bigg)}a_1^\dagger a_2+e^{\sum_{k}\bigg(\frac{g_{1k} }{\omega_k}b_k-\frac{g_{1k}^{\star}}{\omega_k }b_k^\dagger\bigg)}e^{-\sum_{k}\bigg(\frac{g_{2k} }{\omega_k}b_k-\frac{g_{2k}^{\star}}{\omega_k }b_k^\dagger\bigg)}a_2^\dagger a_1 \bigg]\nonumber \\
		&+&\epsilon (n_1+n_2)+\sum_{k} \omega_k b_k b_k^\dagger-\bigg[\sum_{k} n_1^2 \frac{|g_{1k}|^2}{\omega_k}+\sum_{k}n_2^2 \frac{|g_{2k}|^2}{\omega_k}+V_{12}n_1 n_2\bigg]
	\end{eqnarray}
	We assume $g_{{ p}k}=g_k e^{-ik.r_{ p}}$, where $r_{ p}$ is the position vector of the ${ p}$th site. Also, using the identity $e^{X+Y}= e^X e^Y e^{-\frac{1}{2}[X,Y]}$, we simplify above  equation \ref{TEQ} to the following form:
	
	\begin{eqnarray}
		H'&=&J\bigg[e^{-\sum_{k}\alpha_k^{\star}b_k^\dagger}e^{\sum_{k}\alpha_k b_k}e^{-\frac{1}{2}\sum_k |\alpha_k|^2}a_1^\dagger a_2 +e^{-\sum_{k}\alpha_k^{\star}b_k^\dagger}e^{\sum_{k}\alpha_k b_k}e^{-\frac{1}{2}\sum_k |\alpha_k|^2}a_2^\dagger a_1\bigg] \nonumber\\
		&&+\epsilon (n_1+n_2)+  (n_1^2+n_2^2) \sum_{k}\frac{|g_{k}|^2}{\omega_k}+\sum_{k} \omega_k b_k^\dagger b_k+V_{12}n_1 n_2 \nonumber \\ 
		&=& H_S^{\prime} + H_B^{\prime} + H_I^{\prime}		
	\end{eqnarray}
	where 
	\begin{eqnarray}
		H'_S= \tilde{J}[a_1^\dagger a_2+a_2^\dagger a_1]+ \epsilon(n_1+n_2)+V_{12}n_1n_2
	\end{eqnarray}
	represents the system Hamiltonian. Here we have assumed $\frac{|g_k|^2}{\omega_k}<<1$.
	The bath Hamiltonian is given by $H_B= \sum_k\omega_k b_k^\dagger b_k $, and the interaction Hamiltonian in the polaron frame is given by
	\begin{eqnarray}
		H_I^{\prime}=	\tilde{J}[\mathcal{B} a_1^\dagger a_2+\mathcal{B}^\dagger a_2^\dagger a_1].
	\end{eqnarray}
	Here, $	\tilde{J}=Je^{-\frac{1}{2}\sum_k |\alpha_k|^2}
	=Je^{-\frac{1}{2}\sum_k \frac{|g_{1k}-g_{2k}|^2}{\omega_k^2}}$ is the effective hopping energy while
	$\mathcal{B}=e^{\sum_{k}\alpha_k^{\star}b_k^\dagger}e^{{  -}\sum_{k}\alpha_k b_k}-1$ represents the effective bath operators. 
	
	\section{Master Equation}

	The Quantum master is equation given by:
\begin{eqnarray}
	\label{FME}
	\frac{d \rho_s(t)}{dt} =-i{\rm Tr_B}[H'_I(t),\rho_s(0) \rho_B]-\int_0^t {\rm Tr_B}[	H_I^{\prime}(t),[	H_I^{\prime}(\tau),\rho_s(t)\rho_B]],
\end{eqnarray}
{ Since, in the original frame, it is safe to assume the initially uncorrelated qubit-bath state. After, LF transformation, initial system-bath can transform into a coherent state. However, under the approximation anti-adiabatic approximation $\frac{\tilde{J}}{\Delta E_B}<<1$ in polaron frame, it can be shown the leading order corrections are very small and hence an effective description can be done with separable system-bath state. Since, we are at second order of perturbation, it is sufficient to look at first order correction to the wave-function which is given by
\begin{eqnarray}
	|\psi_n\rangle= |\psi^0_m\rangle + \sum_{n\ne m} \frac{\langle \psi^0_n|H_I|\psi^0_m\rangle}{E^0_m-E^0_n}|\psi^0_n\rangle,
\end{eqnarray}
 where $|\psi^0_n$ represent the unperturbed states with unperturbed energy $E_n^0$. Now we assume $|\psi_n^0\rangle= |S\rangle \otimes|0\rangle$ with $H_I =	H_I^{\prime}=	\tilde{J}[\mathcal{B}a_1^\dagger a_2+\mathcal{B}^\dagger a_2^\dagger a_1]$, we see that order of correction term is $\frac{\tilde{J}}{\Delta E_B}$, which by definition is smaller than 1. Therefore, up to second order perturbation, the system-bath remains a separable state.}

For our system in the polaron frame, we have ${\rm Tr_B}[H'_I(t),\rho_s(0) \rho_B]=0$ at $0K$
Therefore we write,
\begin{eqnarray}
	\label{NME}
	\frac{d{\rho_s}(t)}{dt}& = &- \int_0^t d \tau {\rm Tr_B}\big[  {H'}_I(t) {H'}_I(\tau) {\rho}_s(t) \rho_B +{\rho}_s(t) \rho_B {H'}_I(\tau) {H'}_I(t) \nonumber \\&&~~~~~~~~~~~-{H'}_I(\tau){\rho}_s (t) \rho_B{H'}_I(t)-{H'}_I(t){\rho}_s (t)  \rho_B{H'}_I(\tau)\big]
\end{eqnarray}
We calculate the first term of the above equation \ref{NME} , and other terms can be similarly calculated. Using the approximation that system operators does not evolve with time ,we get the interaction Hamiltonian as $H_I^{\prime}(t)=\tilde{J}[B(t)(a_1^\dagger a_2)+B^\dagger(t) (a_2^\dagger a_1)]$,
where
\begin{eqnarray*}
	\mathcal{B}(t)&=&e^{-iH_B't}\mathcal{B}e^{iH_B't}
	\\&=&e^{-iH_B't }\big[e^{\sum_{k}\alpha_k^*b_k^\dagger} e^{-\sum_{k}\alpha_k b_k}-1\big] e^{iH_B' t}
	\\&=&e^{-iH_B't}e^{\sum_{k}\alpha_k^{*}b_k^{\dagger}}e^{iH_B't}e^{-iH_B't}e^{-\sum_{k}\alpha_k b_k}e^{iH_B't}-1
	\\&=&e^{\sum_{k}\alpha_k^{*}b_k^{\dagger}e^{i\omega_{k}t}}e^{-\sum_{k}\alpha_k b_k e^{-i\omega_{k}t}}-1
\end{eqnarray*}
Next we have the first term of equation \ref{NME}; 
\begin{eqnarray}
	Tr_B\big[  {H'}_I(t) {H'}_I(\tau) {\rho}_s(t) \rho_B] &=&\langle 0|[\mathcal{B}(t)a_1^\dagger a_2+\mathcal{B}^{\dagger}(t)a_2^\dagger a_1][\mathcal{B}(\tau)a_1^\dagger a_2+\mathcal{B}^{\dagger}(\tau)a_2^\dagger a_1]  \rho_s(t) |0\rangle \nonumber\\
		&=&\langle 0|[\mathcal{B}(t)\mathcal{B}(\tau)a_1^\dagger a_2 a_1^\dagger a_2]\rho_s(t)|0\rangle+\langle 0|[\mathcal{B}(t)\mathcal{B}^{\dagger}(\tau)a_1^\dagger a_2 a_2^\dagger a_1]\rho_s(t)|0\rangle\nonumber\\
	&&+\langle 0|[\mathcal{B}^{\dagger}(t)\mathcal{B}(\tau)a_2^\dagger a_1a_1^\dagger a_2]\rho_s(t)|0\rangle+\langle 0|[\mathcal{B}^{\dagger}(t)\mathcal{B}^{\dagger}(\tau)a_2^\dagger a_1a_2^\dagger a_1]\rho_s(t)|0\rangle \nonumber
\end{eqnarray}
We calculate these all terms separately as
\begin{eqnarray}
	\langle 0|\mathcal{B}(t)\mathcal{B}(\tau)  |0\rangle&=& \langle 0|(e^{\sum_{k}\alpha_k^{*}b_k^{\dagger}e^{i\omega_{k}t}}e^{-\sum_{k}\alpha_kb_k e^{-i\omega_{k}t}}-1)(e^{\sum_{k}\alpha_k^{*}b_k^{\dagger}e^{i\omega_{k}\tau}}e^{-\sum_{k}\alpha_k b_ke^{-i\omega_{k}\tau}}-1) |0\rangle  \nonumber \\ &=&\langle0|e^{-\sum_{k}\alpha_kb_k e^{-i\omega_{k}t}}e^{\sum_{k}\alpha_k^{*}b_k^{\dagger} e^{i\omega_{k}\tau}}-1|0\rangle \nonumber \\&=&\langle0|e^{\sum_{k}\alpha_k^{*}b_k^{\dagger} e^{i\omega_{k}\tau}}e^{-\sum_{k}\alpha_k^{*}b_k^{\dagger} e^{i\omega_{k}\tau}}e^{-\sum_{k}\alpha_kb_k e^{-i\omega_{k}t}}e^{\sum_{k}\alpha_k^{*}b_k^{\dagger} e^{i\omega_{k}\tau}}-1|0\rangle \nonumber \\&=&\langle0|e^{\sum_{k}\alpha_k^{*}b_k^{\dagger} e^{i\omega_{k}\tau}}e^{-\sum_{k}\alpha_k(b_k +\sum_k \alpha_{k}^{*}e^{i\omega_{k}\tau} )e^{-i\omega_{k}t}}-1|0\rangle\nonumber \\&=&\langle0| e^{\sum_{k}\alpha_k^{*}b_k^{\dagger} e^{i\omega_{k}\tau}}e^{-\sum_{k}\alpha_k b_ke^{-i\omega_{k}t}}e^{-\sum_{k}|\alpha_k|^2e^{-i\omega_k(t-\tau)}}-1|0\rangle\nonumber \\&=&e^{-\sum_{k}|\alpha_k|^2e^{-i\omega_k(t-\tau)}}-1 \nonumber
\end{eqnarray}
And by the similar calculations, we write:
\begin{eqnarray}
	&&	\langle 0|\mathcal{B}(t)\mathcal{B}^{\dagger} (\tau) |0\rangle=e^{\sum_{k}|\alpha_k|^2e^{-i\omega_k(t-\tau)}}-1
	\nonumber \\&&\langle 0|\mathcal{B}^{\dagger}(t)\mathcal{B} (\tau) |0\rangle=e^{\sum_{k}|\alpha_k|^2e^{-i\omega_k(t-\tau)}}-1
\nonumber	\\&&\langle 0|\mathcal{B}^{\dagger}(t)\mathcal{B}^{\dagger} (\tau)|0\rangle=e^{-\sum_{k}|\alpha_k|^2e^{-i\omega_k(t-\tau)}}-1 \nonumber
\end{eqnarray}
Therefore, the final master equation \ref{FME} is given as follows:
\begin{eqnarray}
		\label{FE}
		\frac{d{\rho_s}(t)}{dt} &=& -\tilde{J}^2 \int_0^t d \tau \bigg[\big\{(e^{\sum_{k}|\alpha_k|^2e^{-i\omega_k(t-\tau)}}-1)a_1^{\dagger}a_2a_2^{\dagger}a_1\rho_s+(e^{\sum_{k}|\alpha_k|^2e^{-i\omega_k(t-\tau)}}-1)a_2^{\dagger}a_1 a_1^{\dagger}a_2\rho_s\big\}\nonumber\\
		&&+\big\{(e^{\sum_{k}|\alpha_k|^2e^{i\omega_k(t-\tau)}}-1)\rho_s a_1^{\dagger}a_2a_2^{\dagger}a_1 +(e^{\sum_{k}|\alpha_k|^2e^{i\omega_k(t-\tau)}}-1)\rho_s a_2^{\dagger}a_1 a_1^{\dagger}a_2\}\nonumber\\
		&&-\big\{(e^{-\sum_{k}|\alpha_k|^2e^{i\omega_k(t-\tau)}}-1)a_1^{\dagger}a_2\rho_sa_1^{\dagger}a_2+(e^{\sum_{k}|\alpha_k|^2e^{i\omega_k(t-\tau)}}-1)a_1^{\dagger}a_2\rho_sa_2^{\dagger}a_1\nonumber\\ 
		&&+(e^{\sum_{k}|\alpha_k|^2e^{i\omega_k(t-\tau)}}-1)a_2^{\dagger}a_1\rho_s a_1^{\dagger}a_2+(e^{-\sum_{k}|\alpha_k|^2e^{i\omega_k(t-\tau)}}-1)a_2^{\dagger}a_1\rho_s a_2^{\dagger}a_1\big\}\nonumber\\
		&&-\big\{(e^{-\sum_{k}|\alpha_k|^2e^{-i\omega_k(t-\tau)}}-1)a_1^{\dagger}a_2\rho_sa_1^{\dagger}a_2+(e^{\sum_{k}|\alpha_k|^2e^{-i\omega_k(t-\tau)}}-1)a_1^{\dagger}a_2\rho_sa_2^{\dagger}a_1\nonumber \\
		&&+(e^{\sum_{k}|\alpha_k|^2e^{-i\omega_k(t-\tau)}}-1)a_2^{\dagger}a_1\rho_s a_1^{\dagger}a_2+(e^{-\sum_{k}|\alpha_k|^2e^{-i\omega_k(t-\tau)}}-1)a_2^{\dagger}a_1\rho_s a_2^{\dagger}a_1\big\}\bigg].
\end{eqnarray}
This equation \ref{FE} can be written as;
\begin{eqnarray}
		\frac{d{\rho_s}(t)}{dt} &=&{ i \beta(t)[n_1(1-n_2)+ n_2(1-n_1),\rho_s]} \nonumber \\&&+ \gamma_+(t) \bigg[\big\{a_2^{\dagger}a_1\rho_s a_1^{\dagger}a_2-\frac{1 }{2}\rho_s a_1^{\dagger}a_2a_2^{\dagger}a_1-\frac{1 }{2}a_1^{\dagger}a_2a_2^{\dagger}a_1\rho_s \big\}\bigg] \nonumber\\&&+\gamma_+(t) \bigg[\big\{a_1^{\dagger}a_2\rho_s a_2^{\dagger}a_1- \frac{1 }{2}\rho_s a_2^{\dagger}a_1a_1^{\dagger}a_2-\frac{1 }{2}a_2^{\dagger}a_1a_1^{\dagger}a_2\rho_s \big\}\bigg] \nonumber\\&&+\gamma_-(t)\bigg[a_1^{\dagger}a_2\rho_s a_1^{\dagger}a_2+a_2^{\dagger}a_1\rho_s a_2^{\dagger}a_1\bigg]  
\end{eqnarray}
Where 
	\begin{eqnarray}
		&&\gamma_{\pm}(t)= 2\tilde{J}^2 \int_0^t d\tau \Bigg[ e^{\pm \sum_k |g_{1k}-g_{2k}|^2 \cos [\omega_k (t-\tau)] } \cos \Big(\sum_k |g_{1k}-g_{2k}|^2 \sin [\omega_k (t-\tau)]\Big) -1 \Bigg],\nonumber
		\\&& { \beta(t)=2\tilde{J}^2 \int_0^t d\tau \Bigg[ e^{ \sum_k |g_{1k}-g_{2k}|^2 \cos [\omega_k (t-\tau)] }\sin \Big(\sum_k |g_{1k}-g_{2k}|^2 \sin [\omega_k (t-\tau)]\Big)\Bigg]}.
	\end{eqnarray}
In the singlet-triplet basis, the contribution from first term on r.h.s of the above equation vanishes, due to the fact $n_1(1-n_2)|S/T\rangle=0$. Therefore, we do not consider this terms in the main text.

\end{document}